\documentclass[aps,prl,floatfix,twocolumn,superscriptaddress]{revtex4-2}

\usepackage{graphicx,
            caption,
            subcaption,
            amsmath,
            braket,
            ragged2e,
            xcolor,
            xspace,
            nicefrac,
            lipsum,
            nameref,
            textcomp,
            urwchancal,
            float,
            upgreek,
            isotope,
            hyperref,
            cleveref,
            siunitx,
            placeins, 
            scrextend, 
            url}

\DeclareCaptionLabelSeparator{dot}{. }
\makeatletter
\def\justified{
	\let\\\@normalcr
	\@rightskip\z@skip \rightskip\@rightskip
	\leftskip\z@skip
	\parindent 0em\relax
	\setlength{\parfillskip}{0pt plus 1fil}}
\DeclareCaptionJustification{justified}{\justified}
\hypersetup{colorlinks=true, linkcolor=blue,citecolor=blue,urlcolor=blue}
\def\unit #1 #2 {\SI{#1}{#2}\xspace}
\def\range #1 #2 #3 {\SIrange{#1}{#2}{#3}\xspace}
\DeclareSIUnit\gauss{G}

\newcommand{\myref}[2][]{Fig.~\hyperref[#2]{\ref*{#2}#1}}
\newcommand{\Myref}[2][]{Figure~\hyperref[#2]{\ref*{#2}#1}}
\newcommand{\Mytabref}[2][]{Table~\hyperref[#2]{\ref*{#2}#1}}

\usepackage{bm}

\begin{document}

\title{Study of the inter-species interactions in an ultracold dipolar mixture}

\date{\today}

 
\author{C.~Politi}
\affiliation{
    Institut f\"{u}r Quantenoptik und Quanteninformation, \"Osterreichische Akademie der Wissenschaften, Innsbruck, Austria
}
\affiliation{
    Institut f\"{u}r Experimentalphysik, Universit\"{a}t Innsbruck, Austria
}

\author{A.~Trautmann}
\thanks{Present address: Physikalisches Institut, Auf der Morgenstelle 14 (D-Bau), 72076 Tübingen, Germany}
\affiliation{
    Institut f\"{u}r Quantenoptik und Quanteninformation, \"Osterreichische Akademie der Wissenschaften, Innsbruck, Austria
}

\author{P. Ilzh\"ofer}
\thanks{Present address: 5. Physikalisches Institut and Center for Integrated
Quantum Science and Technology, Universit\"{a}t Stuttgart,
Pfaffenwaldring 57, 70569 Stuttgart, Germany}
\affiliation{
    Institut f\"{u}r Quantenoptik und Quanteninformation, \"Osterreichische Akademie der Wissenschaften, Innsbruck, Austria
}
\affiliation{
    Institut f\"{u}r Experimentalphysik, Universit\"{a}t Innsbruck, Austria
}

\author{G.~Durastante}
\affiliation{
    Institut f\"{u}r Quantenoptik und Quanteninformation, \"Osterreichische Akademie der Wissenschaften, Innsbruck, Austria
}
\affiliation{
    Institut f\"{u}r Experimentalphysik, Universit\"{a}t Innsbruck, Austria
}

\author{M.~J.~Mark}
\affiliation{
    Institut f\"{u}r Quantenoptik und Quanteninformation, \"Osterreichische Akademie der Wissenschaften, Innsbruck, Austria
}
\affiliation{
    Institut f\"{u}r Experimentalphysik, Universit\"{a}t Innsbruck, Austria
}

 \author{M.~Modugno}
 \affiliation{Department of Physics, University of the Basque Country UPV/EHU, 48080 Bilbao, Spain}
 \affiliation{IKERBASQUE, Basque Foundation for Science, 48013 Bilbao, Spain}
	
\author{F.~Ferlaino}
\thanks{Correspondence should be addressed to \mbox{\url{Francesca.Ferlaino@uibk.ac.at}}}
\affiliation{
    Institut f\"{u}r Quantenoptik und Quanteninformation, \"Osterreichische Akademie der Wissenschaften, Innsbruck, Austria
}
\affiliation{
    Institut f\"{u}r Experimentalphysik, Universit\"{a}t Innsbruck, Austria
}


\begin{abstract}
We experimentally and theoretically investigate the influence of the dipole-dipole interactions (DDIs) on the total inter-species interaction in an erbium-dysprosium mixture. By rotating the dipole orientation we are able to tune the effect of the long-range and anisotropic DDI, and therefore the in-trap clouds displacement. We present a theoretical description for our binary system based on an extended Gross-Pitaevskii (eGP) theory, including the single-species beyond mean-field terms, and we predict a lower and an upper bound for the inter-species scattering length $a_{12}$. Our work is a first step towards the investigation of the experimentally unexplored dipolar miscibility-immiscibility phase diagram and the realization of quantum droplets and supersolid states with heteronuclear dipolar mixtures.
\end{abstract}

\maketitle

\section{Introduction}

The ability to tune the interparticle interactions, the geometry and dimensionality of the system, and the possibility of adding complexity in a controlled manner, has made ultracold atomic gases a great platform for studying a plethora of physical phenomena that would be otherwise hard to achieve~\cite{Bloch:2008rev}. Combining two atomic species gives even further opportunities for investigating the effects arising from the interplay between the intra- and inter-species interactions, as polarons~\cite{hu2016bose,jorgensen2016observation}, heteronuclear quantum droplets~\cite{Petrov2015,Cabrera2018,derrico2019}, solitons~\cite{DeSalvo2019}, and ultracold molecules~\cite{Voges2020}.

Heteronuclear mixtures are typically realized by combining contact-interacting atomic species (see alkali-alkali mixtures e.g.~\cite{Modugno2001,Hadzibabic2002,Mudrich2002,Silber2005,McCarron2011,Park2012,Wacker2015}). 
Recently, experiments were able to produce novel types of ultracold mixtures where either one or both mixture components are long-range interacting (lanthanide) atomic species~\cite{Ravensbergen:2018,Trautmann2018}. In particular, the realization of \isotope{Er}-\isotope{Dy} dipolar quantum mixtures is attracting great interest, driven by the possibility of creating new quantum phases even more exotic than the one achieved in contact-interacting mixtures~\cite{Bloch:2008rev} or in single-species dipolar gases~\cite{Norcia2021nof}. Several theory works reported on the study of miscibility in dipolar condensates~\cite{Wilson2012,Kumar2017,Xi2018,Kumar2019}, vortex lattice formation~\cite{Kumar2017vortex,Kumar2018vortex} and on binary quantum droplets realized with dipolar mixtures~\cite{Bisset:2021,Smith:2021,Smith2021approximatetheory}.

In heteronuclear Bose-Bose mixtures, the phenomena mentioned above rely quite strongly on the miscibility-immiscibility conditions. These conditions define whether the two components mix together with the center of masses overlapping at the trap center or whether they are in a phase-separated state where the two center of masses are pushed away from each other. The miscibility-immiscibility phase diagram depends on the contact intra-species scattering lengths, $a_{11}$, $a_{22}$, and dipolar lengths, $a^1_{\text{dd}}$, $a^2_{\text{dd}}$ and the inter-species scattering lengths $a_{12}$ and dipolar lengths $a^{12}_{\text{dd}}$. While $a^{12}_{\text{dd}}$ can be calculated analytically, $a_{12}$ is unknown and its determination relies on experimental measurements.


In this work, we prepare ultracold degenerate mixtures of erbium and dysprosium, and experimentally investigate the effect of the mean-field dipole-dipole interactions on the total inter-species interaction by tracing the center-of-mass displacement for different dipole orientations. We present a theoretical description for our system, including the single-species beyond mean-field terms, which reproduces qualitatively well the experiment. By matching theory and experiment, we define a lower and upper bound for the inter-species scattering length $a_{12}$. 

 \section{Theory}
 Here we consider a binary mixture of dipolar condensates of $^{164}$Dy and $^{166}$Er atoms confined in a harmonic potential, in the presence of a magnetic field $\bm{B}$ aligned along an arbitrary direction in space. The system can be described
 in terms of an extended Gross-Pitaevskii energy functional
 $E = E_{\text{MF}} + E_{\text{dd}} + E_{\text{LHY}}$ with
\begin{align}
E_{\text{MF}} &= 
 \sum_{i=1}^{2}\int \left[\frac{\hbar^2}{2m_{i}}|\nabla \psi_{i}(\bm{r})|^2  + V_{i}(\bm{r})|\psi_{i}(\bm{r})|^{2}\right]\text{d}\bm{r}
 \nonumber\\
 &
+\sum_{i,j=1}^{2}\frac{g_{ij}}{2}\int n_{i}(\bm{r})n_{j}(\bm{r})\text{d}\bm{r},
\end{align}
\begin{equation}
E_{\text{dd}} =\sum_{i,j=1}^{2}\frac{C_{ij}^{\text{dd}}}{2}\iint n_{i}(\bm{r})V_{\text{dd}}(\bm{r}-\bm{r}')n_{j}(\bm{r}') \text{d}\bm{r}\text{d}\bm{r}',
\label{eq:Edd}
\end{equation}
and the single-species Lee-Huang-Yang (LHY) correction for the two components
\begin{equation}
E_{\text{LHY}} = \frac{256\sqrt{\pi}}{15}\sum_{i=1}^{2}\frac{\hbar^{2}a_{ii}^{5/2}}{m_{i}}
\left(1 + \frac{3}{2}\epsilon_{\text{dd},i}^{2}\right)\int n_{i}(\bm{r})^{5/2}\text{d}\bm{r},
\label{eq:ELHY}
\end{equation}
where  $n_i(\bm{r})=|\psi_{i}(\bm{r})|^2$ represents the density of each condensate, $V_{i}(\bm{r})=(m_{i}/2)\sum_{\alpha=x,y,z}\omega_{\alpha,i}^{2}r_{\alpha}^{2} +m_{i}g z$ includes the harmonic trapping and gravity potentials, $g_{ij}=2\pi\hbar^2 a_{ij} (m_{i}+m_{j})/(m_{i}m_{j})$ is the contact interaction strength,
$V_{\text{dd}}(\bm{r})= (1-3\cos^{2}\theta)/(4\pi r^{3})$ the (bare) dipole-dipole potential,
$C_{ij}^{\text{dd}}\equiv\mu_{0}d_{i}d_{j}$ its strength, $d_{i}$ the modulus of the dipole moment $\bm{d}_{i}$ of each species, $\bm{r}$ the distance between the dipoles, and $\theta$ the angle between the vector $\bm{r}$ and the dipole axis, $\cos\theta=\bm{d}\cdot\bm{r}/(dr)$ \cite{Ronen:2006}. In the following we identify species $1$ with the Er condensate, and species $2$ with the Dy condensate (we have also omitted the reference to the mass number, for easiness of notations).
As described later on, the orientation of the magnetic dipoles is varied along arbitrary directions through an external magnetic field $\bm{B}$.

Then, for each set of parameters  the ground state of the system is obtained by minimizing the energy functional $E[\psi_1,\psi_2]$ by means of a conjugate algorithm, see e.\,g.\,Refs.\,\cite{press2007,modugno:2003,Ronen:2006}. In the numerical code the double integral appearing in Eq.\,(\ref{eq:Edd}) is mapped into  Fourier space where it can be conveniently computed using fast Fourier transform (FFT) algorithms, after regularization~\cite{supmat}.
The LHY correction in Eq.\,(\ref{eq:ELHY}) is obtained from the expression for homogeneous 3D dipolar condensates under the local-density approximation \cite{Wachtler:2016,Schmitt:2016}. 

Finally, we remark that the intra-species scattering lengths are given as input to the theory and, although the value for \isotope{Er} has been measured accurately to be $a_{\text{11}} = 83(3)~a_0$ at 
the magnetic field we are working at~\cite{Chomaz:2016}, the intra-species scattering length for \isotope{Dy},  $a_{\text{22}}$, still lacks an accurate determination. Several works have reported different values ranging from $60~a_0$ to $100~a_0$ \cite{Tang2015,Igor2018}. As in these experiments no signs of supersolid or droplet states have been observed~\cite{Chomaz:2019}, we set $a_{\text{22}} = 95~a_0$ for which the ground state is an unmodulated BEC, at our atom numbers and trap frequencies. Since $a_{\text{11}} > a^1_{\text{dd}}$ while $a_{\text{22}} \leq a^2_{\text{dd}}$, with $a^i_{\text{dd}} = C_{ii}^{\text{dd}}m_{i}/(12\pi\hbar^2)$ the dipolar length, we expect Eq.~(\ref{eq:ELHY}) to be more relevant for \isotope{Dy} than for \isotope{Er}.


\section{Experiment}
\begin{figure}[ht]
    \centering
	\includegraphics[width=\columnwidth]{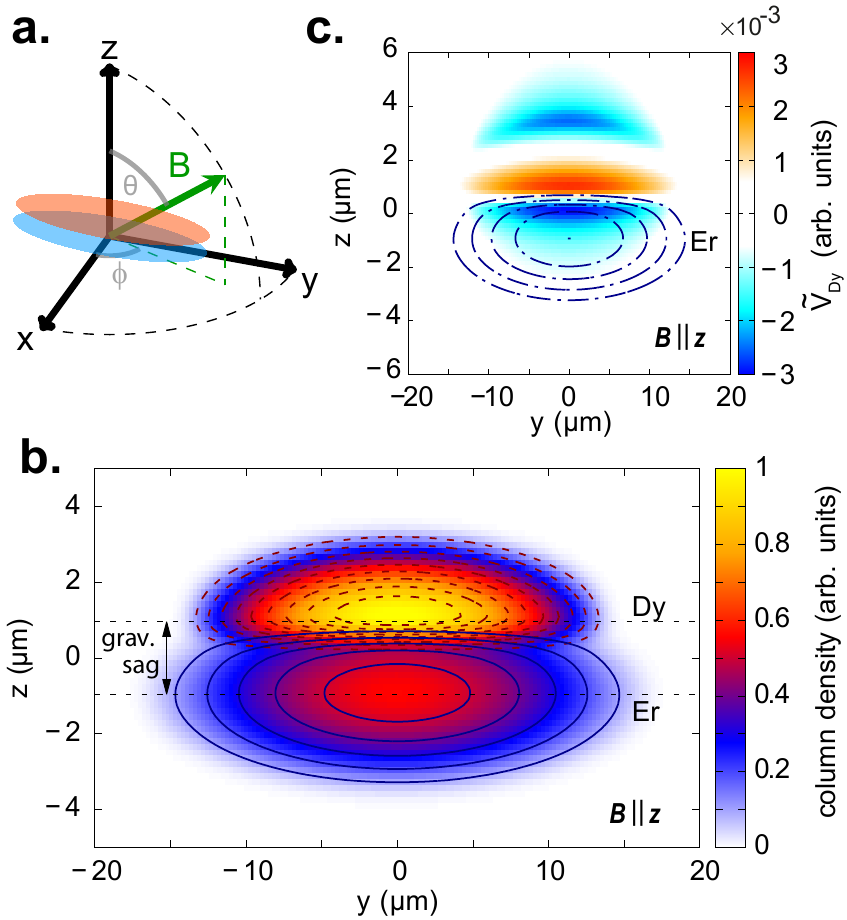}
	\caption {
	{\bf Trap geometry, ground-state column density and dipole potential.}
	\textbf{a.}  
    Illustration of the geometry of our \isotope[164]{Dy} (red ellipse) and \isotope[166]{Er} (blue ellipse) mixture. The orientation of the magnetic field is defined by the angles $\phi$ and $\theta$. The imaging beam propagates in the horizontal plane, at an angle of \ang{45} with respect to the y-axis (not shown). 
    \textbf{b.} Ground-state column density 
    for an imbalanced mixture with $N_{\text{Dy}}=1.3\times10^4$, $N_{\text{Er}}=4.9\times10^4$, $a_{12}=100~a_0$. 
    Dashed and filled lines show the iso-density contour levels for \isotope{Dy} and \isotope{Er}, respectively. For comparison, the in-trap displacement due to the gravitational sag for a non-interacting mixture is also shown (black dashed lines). We set z=0 at the center of the gravitational sag.
    \textbf{c.} Heat map of the dipole potential produced by the Dy condensate (parameters below), $\widetilde{V}_{\isotope{Dy}}(\bm{r})\equiv\int V_{\text{dd}}(\bm{r}-\bm{r}')n_{\isotope{Dy}}(\bm{r}')\text{d}\bm{r}'$, in the $x=0$-plane. Here the magnetic field points along the z-axis. The dotted-dashed lines represent the iso-density contour levels of the \isotope{Er} component, indicating that in this regime the inter-species dipolar interaction is predominantly attractive. 
    	}
	 \label{fig:1} 
\end{figure}
Our experiment starts with a degenerate mixture of \isotope[166]{Er} and \isotope[164]{Dy}, similar to Ref.\,\cite{Trautmann2018}. In brief, after cooling the atomic clouds in a dual-species magneto optical trap~\cite{Ilzhoefer2018}, we start the evaporative cooling by loading the mixture into a single-beam optical dipole trap at $1064$~nm, which propagates horizontally (y-axis); see reference frame in Fig.\,\ref{fig:1}a. After about $600$~ms, the power of a second trapping beam, propagating vertically along the direction of gravity (z-axis), is linearly ramped up to form a crossed optical dipole trap (cODT). Here, the evaporation further proceeds for about $5$~s. We perform the evaporation at a magnetic field of $B = 2.028$~G, pointing along the z-axis, which allows an efficient cooling of both species. 

The final harmonic trap has a cigar-like shape, axially elongated along the y-axis, with frequencies \mbox{$\omega_{x, y, z} = 2 \pi \times (96(1),18(1),150(5))\,\si{\per\second}$}, and \mbox{$\omega_{x, y, z} = 2 \pi \times (104(1),18(1),165(5))\,\si{\per\second}$} for \isotope{Er} and \isotope{Dy}, respectively. 
The trapping frequencies of the two species slightly differ. This is due to the small difference in their mass and atomic polarizability~\cite{Hendrik:2018,Ravensbergen:2018May}.
In a harmonic trap, each species experiences a shift of its center-of-mass (\mbox{COM}) position along the z-axis due to gravity. This effect is known as gravitational sag~\cite{Modugno2003,Hansen2013,Wang2015}. For mixtures, the differential gravitational sag between the components is given by $\Delta z_{\text{grav}}=g(1/\omega_{z1}^2-1/\omega_{z2}^2)$, which for our Er-Dy mixture is $\Delta z_{\text{grav}}=1.9(1) $~µm with Er shifted downwards more than Dy; see Fig.\,\ref{fig:1}a. Such gravitational sag favors  phase separation along the z-axis, reducing the inter-species overlap density. 
In presence of inter-species interactions, the vertical distance of the clouds' centers is not only determined by the gravitational sag but also by their mutual mean-field attraction or repulsion~\cite{Papp2008,McCarron2011,Wacker2015,Lee2016}, quantified by the mean-field shift $\Delta z_{\text{MF}}$. For dipolar mixtures, $\Delta z_{\text{MF}}$ is determined by the interplay between the dipolar and contact inter-species interactions, as we will discuss later.  The total vertical in-trap displacement is thus $\Delta z = \Delta z_{\text{grav}} + \Delta z_{\text{MF}}$.

Figure~\ref{fig:1}b exemplarily shows calculations of the 2D ground-state column density of an imbalanced mixture for  $\bm{B}\parallel \bm{z}$ and $a_{12}=100~a_0$. 
In this configuration, a COM shift is clearly visible, which exceeds the gravitational sag, indicating a total repulsive mean-field interaction between the components. To understand the role of the DDI, it is interesting to calculate the effective potential generated by one species (e.\,g.\,Dy), $\widetilde{V}_{\isotope{Dy}}(\bm{r})\equiv\int V_{\text{dd}}(\bm{r}-\bm{r}')n_{\isotope{Dy}}(\bm{r}')\text{d}\bm{r}'$, felt by the other species (e.\,g.\,Er). Such effective potentials are most relevant in the region where the two species overlap (beside a long-range tail from the DDI). As shown in Fig.\,\ref{fig:1}c, for our trap geometry and dipole orientation, Er experiences a dominant attractive DDI generated by Dy, which is however weaker than the repulsive inter-species contact interaction for $a_{12}=100~a_0$.

\begin{figure}[ht]
    \centering
	\includegraphics[width=\columnwidth]{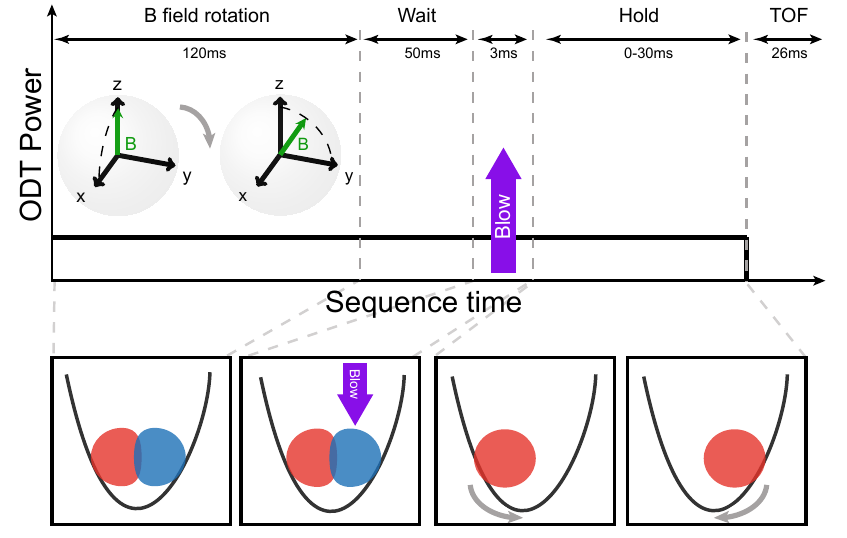}
	\caption {
	{\bf Experimental protocol.} After preparing our Er-Dy mixture with $\bm{B}\parallel \bm{z}$, the magnetic field is rotated to an arbitrary direction, defined by $\theta$ and $\phi$, in $120$~ms. The atomic clouds are held in trap for $50$~ms to reach equilibrium, before either of the species is removed with resonant light. The remaining cloud is held for a variable hold time, $t_h$. The cloud is then released from the trap and imaged with standard absorption imaging after a TOF expansion of $t_{\text{TOF}}=26$~ms. We prepare imbalanced mixtures with condensed atom numbers       $N_{\text{C}}$ in the range $[1-3]\times10^4$ and $[4-6]\times10^4$  for \isotope{Dy} and \isotope{Er}, respectively~\cite{supmat}.
	}
	 \label{fig:2} 
\end{figure}


To experimentally study the inter-species mean-field shift, we selectively remove either one of the two species and follow the COM dynamics of the remaining species towards its new equilibrium position in the trap~\cite{Trautmann2018}.
Figure~\ref{fig:2} illustrates our  protocol. 
After preparing our trapped Bose-Bose Er-Dy mixture with $\bm{B}\parallel \bm{z}$,  we first adiabatically rotate the magnetic field in $120$~ms to the desired orientation (i.\,e.\,changing $\theta$ and $\phi$) and let the mixture equilibrate for $50$~ms. We then selectively remove either \isotope{Er} or \isotope{Dy} by shining a resonant light pulse, operating on either of the two strong atomic transitions ($401$~nm ($421$~nm) for Er (Dy)).
We have checked that this resonant pulse of $3$-ms duration does not affect the remaining species.
Finally, we hold the remaining species in trap for a variable time, $t_h$, and probe the system with standard absorption imaging after a time-of-flight (TOF) expansion of $t_{\text{TOF}}=26$~ms.

After the selective removal of either of the two species, the remaining species is out of equilibrium and the cloud \mbox{COM} starts to oscillate 
around its new equilibrium position, given by the dipole-trap minimum in presence of gravity.
Figure~\ref{fig:3} a(b) exemplarily shows the vertical \mbox{COM} position, $Z_i$~\cite{supmat}, measured after TOF, for Dy (Er) after removing Er (Dy) and for two different dipole orientations.

The amplitude of the observed oscillation is directly connected to the inter-species mean-field shift experienced by the atoms in trap. 
Within the assumption of ballistic expansion, which is justified in the weakly interacting regime,  $Z_i(t_h,t_{\text{TOF}}) =  z_i(t_h) +\dot{z}_i(t_h)t_{\text{TOF}} + gt_{\text{TOF}}^2/2$, where $z_i(t_h) =  \Delta z_{\text{MF},i}\cos(\omega_{i}t_h) + \Delta z_{\text{grav}}$ is the in-trap COM position.
The oscillation frequency $\omega_{i}$ is the trap frequency along the z-axis. 

By combining the previous equations, one gets the following expression
\begin{equation}
\begin{split}
Z_i(t_h,t_{\text{TOF}}) =~& \Delta z_{\text{MF},i}~\cos(\omega_{i}t_h)~ \\
& - \Delta z_{\text{MF},i}~\omega_{i}\sin(\omega_{i}t_h)~t_{\text{TOF}} + z_{\text{off}},  
\end{split}
\label{eq:fitfun}
\end{equation}
where $z_{\text{off}} = \Delta z_{\text{grav}} + gt_{\text{TOF}}^2/2$.
We fit  Eq.\,(\ref{eq:fitfun}) to the experimental data for the two magnetic field orientations with the mean-field shift $\Delta z_{\text{MF},i}$, $\omega_i$, and $z_\text{off}$ being free fitting parameters.

\section{Results}
\begin{figure}[ht]
    \centering
	\includegraphics[width=\columnwidth]{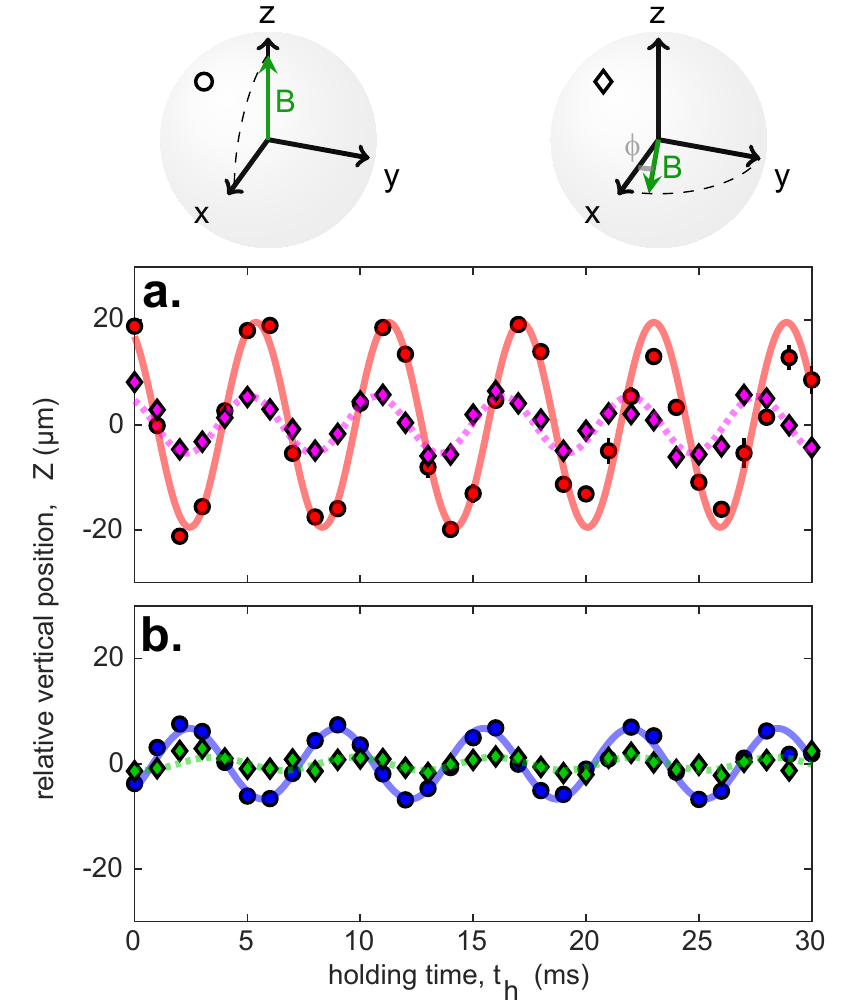}
	\caption {
	{\bf COM oscillations after removal of either one of the species.} \textbf{a.} Vertical COM position of \isotope{Dy} after removing \isotope{Er} and (\textbf{b}) viceversa. The vertical position $Z_i $ is recorded after a TOF expansion of $26$ms, as a function of the holding time. The offset $z_{\text{off}}$ has been subtracted to facilitate comparison. The measurements are repeated for two magnetic field orientations: $\bm{B}\parallel \bm{z}$ \textendash~$\theta=\ang{0}$, $\phi=\ang{0}$~\textendash~(circles) and $\bm{B}\in \bm{xy}$ \textendash~$\theta=\ang{90}$, $\phi=\ang{15}$~\textendash~(diamonds). The atom numbers are $N_{\text{Dy}}=1.3(2)\times10^4$, $N_{\text{Er}}=4.9(7)\times10^4$ and $N_{\text{Dy}}=3.1(5)\times10^4$, $N_{\text{Er}}=4.7(5)\times10^4$ for $\bm{B}\parallel \bm{z}$ and $\bm{B}\in \bm{xy}$, respectively. The error bars reported represent the standard error on the mean over three experimental trials, and are mostly smaller than the markers. We fit Eq.\,(\ref{eq:fitfun}) to the data for $\bm{B}\parallel \bm{z}$ (filled lines) and $\bm{B}\in \bm{xy}$ (dotted lines).
	}
	 \label{fig:3} 
\end{figure}

Figure~\ref{fig:3} shows important information on the inter-species interactions.  First, by comparing the dynamics of the two species, we  observe that the oscillations are counter-phase. The \isotope{Dy} cloud starts moving downwards towards the trap center, whereas the \isotope{Er} one moves upwards, confirming a total repulsive inter-species interaction for this geometry.  Second, we see a clear difference in the oscillation amplitude between \isotope{Dy} and \isotope{Er}. This is due to the fact that the mixture is imbalanced with \isotope{Er} being the majority species, and therefore the mean-field shift caused by \isotope{Er} on \isotope{Dy} is larger. Finally, for each species, the oscillation amplitude strongly depends on the magnetic field orientation. This behaviour can not be simply explained by the anisotropy of the DDI.
For $\bm{B}\parallel \bm{z}$, the DDI is more attractive over the inter-species overlap region than for $\bm{B}$ in the xy-plane, $\bm{B}\in \bm{xy}$. Hence, one would expect \mbox{$\Delta z_{\text{MF,z}}<\Delta z_{\text{MF,xy}}$}, contrasting the observations. 

The additional effect to account for is the magnetostriction~\cite{Santos2005} of each species, i.e.~a cloud elongation  along the magnetization direction caused by the single-species DDI.  
For $\bm{B}\parallel \bm{z}$,  the two clouds elongate along the z-axis, thus increasing the inter-species overlap density; see Fig.\,\ref{fig:1}b. This increased overlap activates a back action on the strength of the repulsive contact interaction, which acquires a larger weight, leading to an increased repulsion between the clouds. On the contrary, for $\bm{B}\in \bm{xy}$, the clouds elongate  horizontally thereby minimizing the overlap density and therefore the inter-species repulsion. The slight difference in frequency observed for the two magnetic-field orientations is due to the presence of small residual magnetic-field gradients~\cite{supmat}. 



To get further insight into the anisotropy of the interspecies interactions, we repeat the above measurement for various dipole orientations, set by  the angles $\theta$ and $\phi$.
As before, we perform two sets of measurements: we probe the out-of-equilibrium dynamics of \isotope{Dy} after removing \isotope{Er} and viceversa. To enhance the amplitude of the COM oscillations of one species (\isotope{Dy}), we perform measurements with imbalanced mixtures, where \isotope{Er} is the majority species with condensed atom numbers in the range $[4-6]\times10^4$,  while the \isotope{Dy} cloud contains about $[1-3]\times10^4$~\cite{supmat}.

\begin{figure}[ht]
    \centering
	\includegraphics[width=\columnwidth]{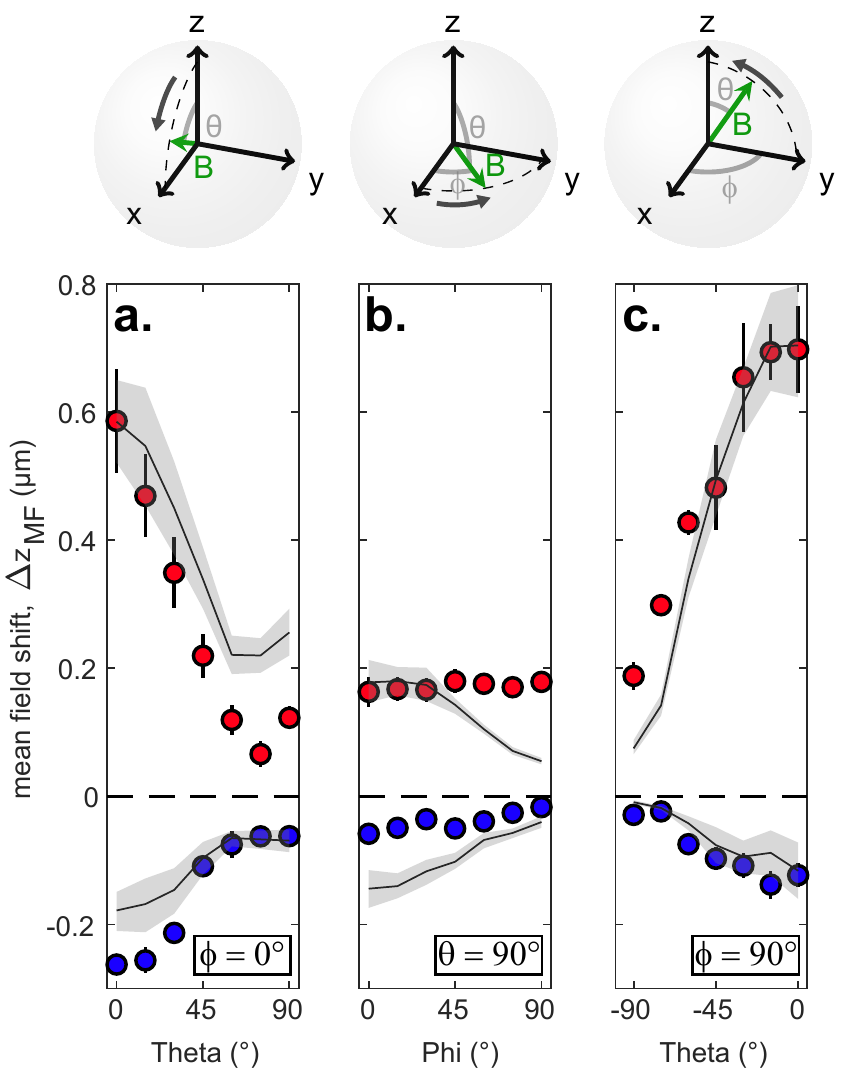}
	\caption {
	{\bf Mean-field displacement and theory prediction.} 
	Experimental estimation of the mean-field displacement $\Delta z_{\text{MF},i}$ for \isotope{Dy} (red points) and \isotope{Er} (blue points), as a function of the magnetic field orientation. \textbf{a.} $\theta=[\ang{0},\ang{90}]$, $\phi=\ang{0}$. \textbf{b.} $\phi=[\ang{0},\ang{90}]$, $\theta=\ang{90}$. \textbf{c.} $\theta=[\ang{0},\ang{90}]$, $\phi=\ang{90}$. Theory prediction for an inter-species scattering length $a_{12}=100~a_0$ (gray lines). The gray shaded area takes into account the experimental uncertainty on the estimation of the atom number. The error bars in $\Delta z_{\text{MF},i}$ correspond to the statistical uncertainty from the fit. The mismatch between the data points at $\theta=\ang{0}$ in \textbf{a.} and \textbf{c.} is due to different atom numbers~\cite{supmat}.
	}
	 \label{fig:4} 
\end{figure}

Figure~\ref{fig:4} summarizes our results. It shows both the measured and calculated mean-field shift $\Delta z_{\text{MF},i}$ for each plane of rotation for \isotope{Dy} (red points) and \isotope{Er} (blue points). 
We observe that $\Delta z_{\text{MF},i}$ has a  maximum for $\bm{B}\parallel \bm{z}$ and decreases when approaching the horizontal plane. The gray lines show the theory results for an inter-species scattering length $a_{12}=100~a_0$ and for our experimental parameters, i.e. atom numbers and trap frequencies. We chose $a_{12}=100~a_0$ as it describes best the experimental data. The gray shaded area takes into account the experimental uncertainty on the estimation of the atom number. 

The theory curves agree qualitatively with the experimental observations. In particular, experiment and theory are in good agreement for $\bm{B}\parallel \bm{z}$, while they start to deviate for $\bm{B}\in \bm{xy}$. The small mismatch can be due to the presence of residual vertical magnetic-field gradients, which are not taken into account in the theory. These can cause a systematic shift of the trap frequencies to higher values when going from  $\bm{B}\parallel \bm{z}$ to $\bm{B}\in \bm{xy}$ thereby reducing the gravitational sag~\cite{supmat}. Furthermore, while our \isotope{Dy} ground-state calculations predict the transition to a macrodroplet at $a_{\text{22}} = 95~a_0$ for $\bm{B}\parallel \bm{y}$, and a further reduction of the overlap density, in the experiment we observe a stable \isotope{Dy} BEC. Previous works have also shown a quantitative mismatch between theory and experiment in predicting the macrodroplet transition, suggesting the need of refined models and an accurate determination of $a_{\text{22}}$~\cite{Schmitt:2016,Chomaz:2016,Bottcher2019b}.



The overall behaviour shown in Fig.~\ref{fig:4} can be explained by the effect of the magnetostriction on the inter-species overlap density. In fact, as discussed earlier, for magnetic field orientations in the horizontal plane, the clouds are elongated horizontally along the direction of $\bm{B}$ thereby minimizing the density overlap and the inter-species repulsion. 
Whereas, when orienting the magnetic field along the vertical direction, the magnetostriction leads to an increase of the density overlap and therefore of the inter-species repulsion, which overcomes the attractive DDI. The system undergoes a transition to a state where the two components are pushed aside, maximising the in-trap displacement (see Fig.\,\ref{fig:1}b).

\begin{figure}[ht]
   \centering
	\includegraphics[width=0.9\columnwidth]{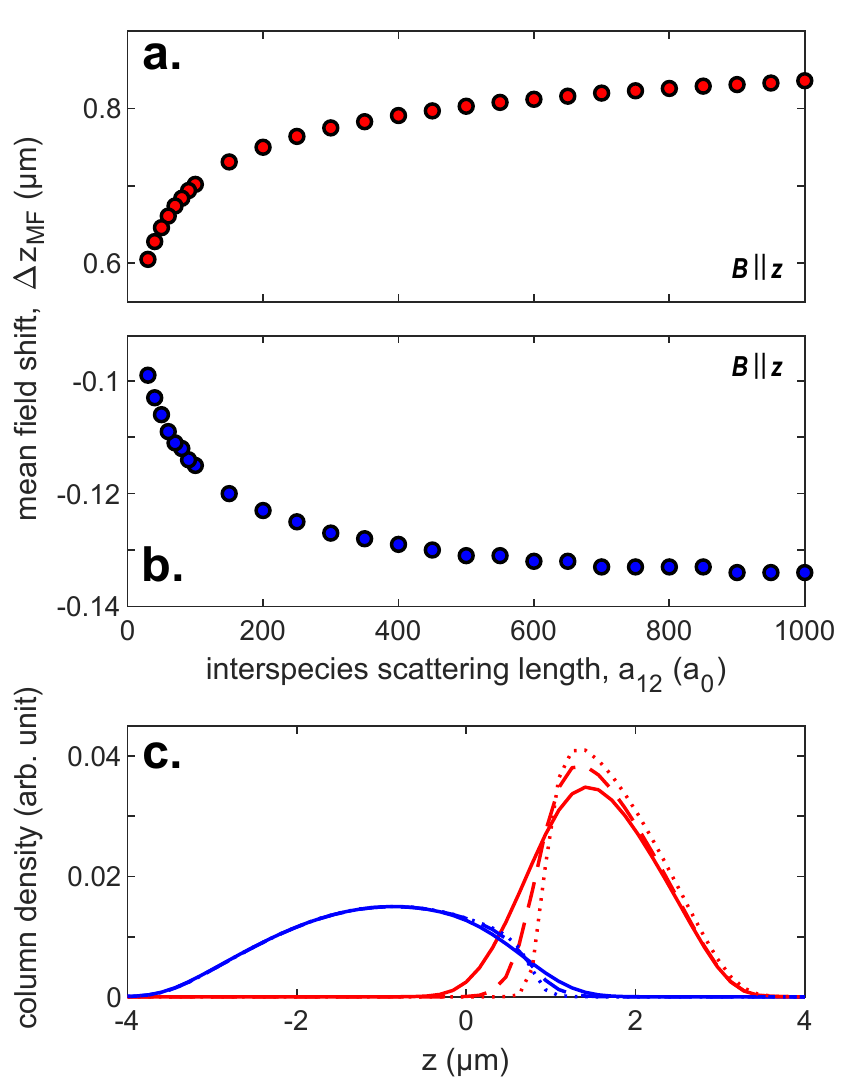}
	\caption {{\bf Calculated mean-field displacement as a function of $a_{12}$.} Calculated mean-field displacement for \isotope{Dy} (\textbf{a.}) and \isotope{Er} (\textbf{b.}) as a function of the inter-species scattering length $a_{12}$. The magnetic field is oriented along the z-axis. \textbf{c.} In-trap density cut along $y=0$ for Dy (red) and Er (blue), for $a_{12}=30~a_0$ (filled lines), $a_{12}=100~a_0$ (dashed lines) and $a_{12}=200~a_0$ (dotted lines). Here $N_{\text{Dy}}=0.8\times10^4$ and $N_{\text{Er}}=5.9\times10^4$.}
	 \label{fig:5} 
\end{figure}

For a magnetic field orientation along the z-axis, we perform ground-state calculations varying the inter-species scattering length $a_{12}$ and calculate the \isotope{Er}-\isotope{Dy} mean-field displacement as a function of $a_{12}$. The results are shown in Fig.\,\ref{fig:5}. The mean-field displacement increases with $a_{12}$ owing to the fact that \isotope{Dy} (\textbf{a.}) is pushed away from \isotope{Er} (\textbf{b.}). Figure\,\ref{fig:5}c shows the \isotope{Dy} (red) and \isotope{Er} (blue) density cuts along $y=0$, for $a_{12}=30~a_0$ (filled lines), $a_{12}=100~a_0$ (dashed lines) and $a_{12}=200~a_0$ (dotted lines). The repulsive interaction between the species leads to a decrease of the density overlap when going to higher $a_{12}$. We compare the theory results with the experimentally measured mean-field shift at $[\theta=\ang{0}$, $\phi=\ang{90}]$, and by performing a \mbox{$\chi^2$-analysis} we are able to estimate the inter-species scattering length to be, $a_{12} = 110~[-70,+240]~a_0$~\cite{supmat}. From our ground-state calculations, when choosing \mbox{$a_{12}<30~a_0$} the repulsive contribution of the contact interactions to the mean-field shift is not enough to overcome the attractive contribution from the DDI (see Fig.\,\ref{fig:1}c) leading to a collapse of both species. In this regime, it might be necessary to include the inter-species LHY term as done in \cite{Bisset:2021,Smith:2021}.



 \section{Conclusions and outlook}
 
\begin{figure}[ht]
    \centering
	\includegraphics[width=\columnwidth]{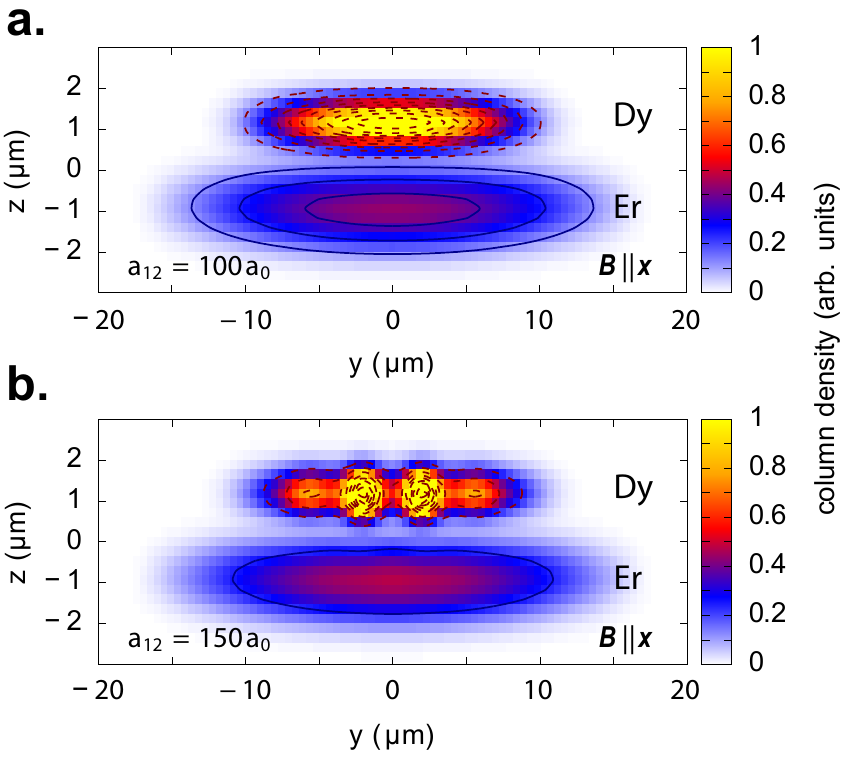}
	\caption{\textbf{Calculation of an interaction-induced supersolidity.} Ground-state configurations for an imbalanced dipolar mixture with the magnetic field pointing along the x-axis, for two different values of the inter-species scattering length: \textbf{a.  $a_{12}=100~a_0$}, \textbf{b.  $a_{12}=150~a_0$}.
  Dashed and filled lines show the iso-density contour levels for \isotope{Dy} and \isotope{Er}, respectively. Here $N_{\text{Dy}}=1.2\times10^4$, $N_{\text{Er}}=6\times10^4$, $a_{11} = 83~a_0$ and $a_{22} = 95~a_0$.
}
	 \label{fig:6} 
\end{figure}

In conclusion, we have experimentally investigated the effect of the DDI on the total inter-species interaction by tracing the mean-field in-trap displacement between the species. We have presented a theoretical description for our \isotope{Er}-\isotope{Dy} mixture, including the single-species beyond mean-field corrections, which qualitatively describes well our system and allows us to predict an inter-species scattering length on the order of $a_{12}=100~a_0$. By changing the magnetic field orientation from the horizontal plane to the vertical direction, we were able to observe a transition to a state in which the two components are pushed apart by the dominant mean-field repulsive interaction. Future studies will focus on the use of inter-species Feshbach resonances, recently reported in our group \cite{Durastante2020}, to reach the conditions in which one or both components exhibit a phase transition to a quantum droplet or supersolid regime. As an example, Fig.~\ref{fig:6} shows that the onset of a supersolid phase in the Dy component can be induced by increasing the inter-species contact scattering length $a_{12}$.

\section{Acknowledgments}
\begin{acknowledgments}
We thank Maximilian Sohmen for the experimental support and for valuable discussions. We thank Rick van Bijnen for theoretical support. We also thank Lauriane Chomaz, Matthew Norcia, Lauritz Klaus and the Innsbruck Erbium team for fruitful discussions.  
This work is financially supported through an ERC Consolidator Grant (RARE, No.\,681432), an NFRI grant (MIRARE, No.\,\"OAW0600) of the Austrian Academy of Science, the QuantERA grant MAQS by the Austrian Science Fund FWF No.\,I4391-N, and the DFG/FWF via FOR~2247/PI2790. M.\,M.~acknowledges support by the Spanish Ministry of Science, Innovation and Universities and the European Regional Development Fund FEDER through Grant No. PGC2018-101355-B-I00 (MCIU/AEI/FEDER, UE), and by the Basque Government through Grant No. IT986-16.
We also acknowledge the Innsbruck Laser Core Facility, financed by the Austrian Federal Ministry of Science, Research and Economy.

\end{acknowledgments}

\bibliography{reference}



\clearpage
\appendix
\renewcommand\thefigure{\thesection S\arabic{figure}}   
\setcounter{figure}{0}   

\section{Supplemental Material}

\subsection{Atom number and vertical COM position}
\begin{figure}
   \centering
	\includegraphics[width=\columnwidth]{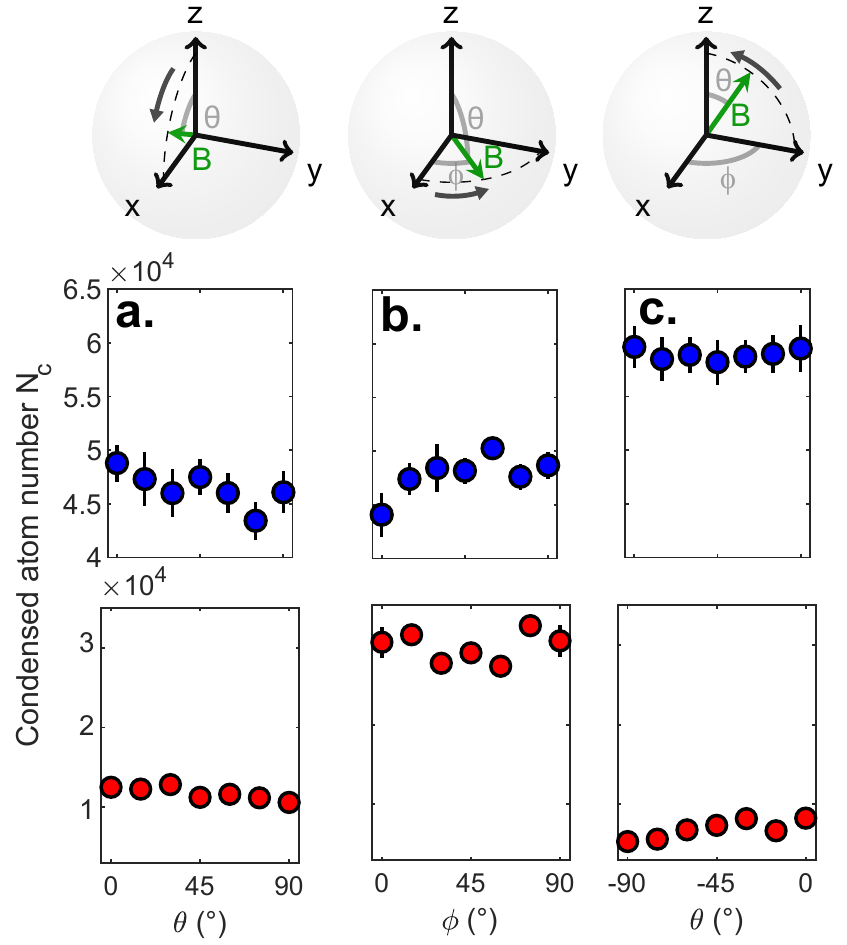}
	\caption {{\bf Condensed atom numbers as a function of the magnetic field orientation.} Measured atom numbers in the BEC for \isotope{Dy} (red points) and \isotope{Er} (blue points) related to the measurement shown in Fig.\,\ref{fig:4} of the main text. The magnetic field is oriented in the XZ-plane (\textbf{a.}), in the XY-plane (\textbf{b.}), and in the YZ plane (\textbf{c.}). The error bars reported represent the standard error on the mean over three experimental trials.}
	 \label{fig:atomnum} 
\end{figure}

After each experimental sequence -- described in Fig.\,\ref{fig:2} -- we release the clouds and perform absorption imaging after a TOF expansion of $26$~ms. We measure the condensed atom number for each species after substracting the thermal part by fitting a symmetric 2D Gaussian to the wings of the density distribution. We then fit an asymmetric Gaussian to the remaining density distribution to extract the vertical COM position $Z_i$. Figure~\ref{fig:atomnum} shows the measured condensed atom numbers $N_{\text{C}}$ of \isotope{Dy} (red points) and \isotope{Er} (blue points), related to the results presented in Fig.\,\ref{fig:4} of the main text. These atom numbers are given as input to the theory for each value of $\theta$ and $\phi$.

\subsection{Dipolar energy: Fourier representation and regularization}

Here we outline the method used for calculating the double integral in Eq.~(\ref{eq:Edd}),
following the standard approach introduced in Ref.~\cite{Ronen:2006}.
As anticipated, we start by rewriting the above integral in Fourier space.
In particular, we make use of the \mbox{\textit{Parseval theorem}~\cite{press2007}}, 
$\int g(\bm{x})h^{*}(\bm{x}) \text{d}\bm{x} = \int \tilde{g}(\bm{k})\tilde{h}^{*}(\bm{k}) \text{d}\bm{k}$,
where $\tilde{g}(\bm{k})\equiv FT[g](\bm{k})$ and $\tilde{h}^{*}(\bm{k})\equiv\left\{ FT[h](\bm{k})\right\}^{*}$. Then, by 
defining $f\equiv h^{*}$ and recalling that $FT[f^{*}](\bm{k})=\tilde{f}^{*}(-\bm{k})$, we have
$\tilde{h}^{*}(\bm{k})=\left\{ FT[f^{*}](\bm{k})\right\}^{*}=\tilde{f}(-\bm{k})$, so that 
\begin{equation}
E_{\text{dd}} = \frac12\int \tilde{n}_i^{*}(\bm{k})\widetilde{V}_{\text{dd}}(\bm{k})\tilde{n}_j(\bm{k}) \text{d}\bm{k},
\end{equation}
where we have used the fact that $n_i(\bm{r})$ is real, which implies $\tilde{n}_i(-\bm{k})=\tilde{n}_i^{*}(\bm{k})$ \footnote{We notice that the expression in Eq. (A8) of Ref.~\cite{Ronen:2006}  can be used only if $n(\bm{r})$ is even under parity in the three spatial directions, namely when $\tilde{n}(-\bm{k})=\tilde{n}(\bm{k})$. }.
At this point it is worth to recall that the use of the FT implicitly entails a periodic system, and this brings along an unwanted effect: the long range dipolar interactions can couple the system to virtual periodic replica. Such a coupling is obviously unphysical, and it can be cured by limiting the range of the dipolar interaction within a sphere of radius $R$ (contained inside the computational box of size $L$), namely multiplying $V_{\text{dd}}(\bm{r})$ by the Heaviside step function $\Theta(R-r)$, with $R\le L/2$. The corresponding FT is \cite{Ronen:2006}
\begin{equation}
\widetilde{V}_{\text{dd}}^{\text{cut}}(\bm{k})=
4\pi\left(1 + 3\frac{\cos(Rk)}{R^{2}k^{2}} -3\frac{\sin(Rk)}{R^{3}k^{3}} \right)
\left(\cos^{2}\alpha -\frac13\right).
\end{equation}


\subsection{Estimation of the residual magnetic field gradient.}
To evaluate the residual magnetic field gradient we measure the COM position of Er and Dy as a function of the TOF and for different values of the magnetic field. In this way, we are able to extract the correction to the gravitational acceleration $g$ due to residual magnetic field gradients. When $\bm{B}$ is oriented along the z-axis we measure an increase in $g$ of about $2\%$ for \isotope{Dy} and $1\%$ for \isotope{Er}. The presence of these residual magnetic field gradients along the direction of gravity leads to a slight decrease of the trap freqencies (see Fig.~\ref{fig:3} in the main text) when orienting $\bm{B}$ from the XY-plane to the z-axis. The tensorial polarizability~\cite{Lepers2017,Hendrik:2018} could also cause a shift of the trap frequencies when changing the orientation of the magnetic field, but these are negligible in our case.


\subsection{Estimation of the inter-species scattering length.}
From the calculated mean-field shift as a function of the inter-species scattering length, shown in Fig.~\ref{fig:5} of the main text, we can estimate the value of $a_{12}$ that best represents the experimental results and its confidence interval. To do that, we perform a $\chi^2$-analysis for the mean-field shift at $[\theta=\ang{0}$, $\phi=\ang{90}]$, with \mbox{$\chi^2 = (\Delta z_{\text{MF}}-\Delta z^{\text{th}}_{\text{MF}})^2/\sigma^2_{\Delta z_{\text{MF}}}$}, where $\Delta z^{\text{th}}_{\text{MF}}$ is the theoretically calculated in-trap mean-field displacement, and $\Delta z_{\text{MF}}$ and $\sigma_{\Delta z_{\text{MF}}}$ the experimental value and its statistical error. By fitting a Gaussian around the minimum of the distribution and by defining its confidence interval as the range in which \mbox{$\chi^2<1$}~\cite{Young:2015}, we estimate $a_{12} = 110~[-70,+240]~a_0$. 

\begin{figure}
   \centering
	\includegraphics[width=\columnwidth]{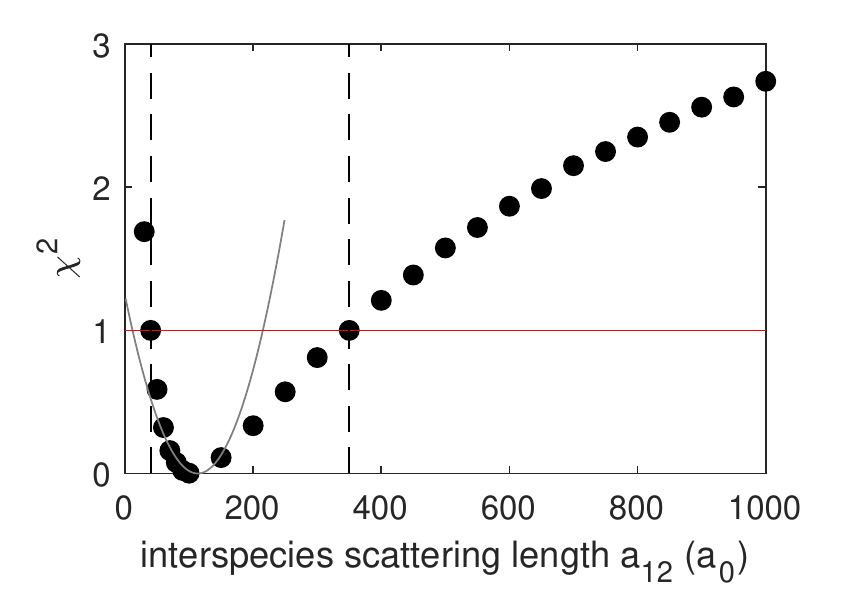}
	\caption {{\bf $\chi^2$-distribution of the mean-field shift.}  \mbox{$\chi^2$-distribution} for our \isotope{Er}-\isotope{Dy} mean-field shift (black points). We estimate the inter-species scattering length to be $a_{12} = 110~[-70,+240]~a_0$, by locally doing a Gaussian fit around the minimum of the distribution (gray line) and by defining the lower and upper bounds as the values at which \mbox{$\chi^2=1$} (black dashed lines).}
	 \label{fig:chi2} 
\end{figure}

\end{document}